\documentclass[letterpaper, 10 pt, conference]{ieeeconf}  %

\IEEEoverridecommandlockouts                              %

\usepackage{color}
\usepackage{enumerate}
\usepackage{graphicx}
\usepackage{varwidth}%
\usepackage{mathrsfs}
\usepackage{mathtools}
\usepackage[font=small,labelfont=bf]{caption}
\usepackage{wrapfig}
\usepackage{overpic}
\usepackage{multirow}
\usepackage[dvipsnames]{xcolor}
\usepackage{braket}

\usepackage{xcolor}
\definecolor{darkred}{RGB}{139,0,0}

\newcommand{\algcomment}[1]{%
  \textbf{\footnotesize\color{black!55}\ttfamily#1}%
}

\allowdisplaybreaks

\usepackage{amsmath,amssymb,colonequals,etoolbox}
\usepackage{thmtools}
\usepackage{url}
\usepackage{algorithm}
\usepackage{algpseudocode}
\algnewcommand{\algorithmicinput}{\textbf{Input:}}
\algnewcommand{\algorithmicoutput}{\textbf{Output:}}
\algnewcommand{\Input}[1]{\Statex \algorithmicinput\ #1}
\algnewcommand{\Output}[1]{\Statex \algorithmicoutput\ #1}

\usepackage{subcaption}
\usepackage{threeparttable}
\usepackage{stfloats}

\usepackage{mdframed}
\usepackage{booktabs}

\usepackage{cite}

\usepackage{hyperref}
\hypersetup{
    colorlinks,
    linkcolor={red!50!black},
    citecolor={red!50!black},
    urlcolor={red!80!black}
}
\usepackage{cleveref}
\crefname{figure}{Fig.}{Figs.}
\Crefname{figure}{Fig.}{Figs.}

\crefname{algorithm}{Alg.}{Algs.}
\Crefname{algorithm}{Alg.}{Algs.}

\crefname{section}{Sec.}{Secs.}
\Crefname{section}{Sec.}{Secs.}

\renewcommand{\geq}{\geqslant}
\renewcommand{\le}{\leqslant}
\renewcommand{\leq}{\leqslant}

\newtheorem{thm}{Theorem}[section]

\newtheorem{myex}[thm]{Exercise}

\mdfdefinestyle{theoremstyle}{%
linecolor=black,linewidth=1pt,%
frametitlerule=true,%
frametitlebackgroundcolor=gray!20,%
innertopmargin=0em,%
innerbottommargin=0.4em,%
nobreak=true,%
}

\DeclareMathOperator*{\argmin}{arg\!\min}
\DeclareMathOperator*{\argmax}{arg\!\max}

\newcommand{\calU}{\mathcal{U}}
\newcommand{\calV}{\mathcal{V}}

\newcommand{\calL}{\mathcal{L}}
\newcommand{\calD}{\mathcal{D}}

\newcommand{\R}{\ensuremath{\mathbb{R}}}

\newcommand{\norm}[1]{\lVert #1 \rVert}

\newcommand{\ip}[2]{\ensuremath{\langle #1, #2 \rangle}}

\newcommand{\E}{\mathbb{E}}
\newcommand{\abs}[1]{\ensuremath{| #1 |}}

\newcommand{\ind}{\mathbf{1}}

\newcommand{\calB}{\mathcal{B}}

\newcommand{\calI}{\mathcal{I}}

\newcommand{\calT}{\mathcal{T}}
\newcommand{\calF}{\mathcal{F}}

\newcommand{\calS}{\mathcal{S}}

\newcommand{\sfU}{\mathsf{U}}
\newcommand{\sfD}{\mathsf{D}}
\DeclarePairedDelimiterX{\infdivx}[2]{(}{)}{%
  #1\;\delimsize\|\;#2%
}

\newcommand{\e}{\varepsilon}

\newcommand{\rmd}{\mathrm{d}}

\newcommand{\mkmat}[1]{\ensuremath{\begin{bmatrix}#1\end{bmatrix}}}

\newcommand{\MethodName}{\textsc{Steer2Reach}}
\newcommand{\MethodAbbrv}{\textsc{S2R}}

\title{\bf
Forward Trajectory Steering for Hamilton-Jacobi Reachability Analysis
}

\author{Sungje Park and Stephen Tu%
\thanks{This work was partially supported by the National Science Foundation
through NSF CPS \#2434460 and NSF GRFP (DGE-1842487). Authors are with the Ming Hsieh Department of Electrical and Computer Engineering, University of Southern California, Los Angeles, CA, USA. Corresponding E-mail: \href{mailto:sungjepa@usc.edu}{sungjepa@usc.edu}.}
}

\begin{document}

\maketitle
\thispagestyle{empty}
\pagestyle{empty}

\begin{abstract}
Hamilton-Jacobi (HJ) reachability provides a mathematically rigorous framework for 
safe control of dynamical systems, 
but its practical application is bottlenecked by the computational complexity of solving Hamilton-Jacobi-Isaacs variational inequality PDEs in high dimensions. 
Physics-informed neural networks (PINNs) have recently emerged as a promising alternative to classical mesh-based solvers, yet their performance is highly sensitive to the choice of collocation sampling.
In order to learn accurate safety value functions, 
existing PINNs-based HJ reachability solvers must rely on complex training pipelines and auxiliary supervision.

In this work, we propose \MethodName{} (\MethodAbbrv{}), a PINNs-based HJ reachability solver that requires minimal modification on top of standard PINNs training. 
\MethodAbbrv{}'s key contribution is a lightweight, low-overhead
adaptive collocation sampling distribution constructed by 
steering forward trajectories using 
a combination of the optimal control and disturbance signals induced by the current value function, with injected stochastic exploration noise.
We demonstrate that despite its simplicity, \MethodAbbrv{} achieves competitive---and in some cases improved---performance on safety metrics while reducing relative $L_2$ error across a range of reachability benchmarks
compared with SoTA MPC-guided HJ reachability solvers, all without requiring multi-stage training or MPC-based supervision.\footnote{Github repo: \href{https://github.com/sungje-park/steer2reach}{github.com/sungje-park/steer2reach}.}

\end{abstract}

\section{Introduction}
\label{sec:intro}

As autonomous systems such as self-driving cars, unmanned aerial vehicles, and robotic assistants become more prevalent in society, the need for formal safety assurances is increasingly crucial.
Hamilton-Jacobi (HJ) reachability analysis~\cite{mitchell2005reachability} is an 
appealing approach for addressing this need, providing an exact mathematical characterization of the safe operation regimes of a control system. 
Specifically, through the computation of a safety value function, HJ reachability identifies a safe controller that ensures the system avoids reaching unsafe states, even subject to adversarial disturbances that may arise during execution.

Traditionally, HJ reachability requires computing a numerical solution to a Hamilton-Jacobi-Isaacs variational inequality (HJI-VI), which describes the safety value function. Classical numerical techniques solve the HJI-VI partial differential equation (PDE) over a dense state space grid, and hence suffer from a curse of dimensionality as the state dimension grows~\cite{bansal2017reachability}. 
This is a significant limitation for modern control systems---e.g., high DoF articulated manipulators, humanoid robots, and multi-agent systems---which often involve high-dimensional states. 
In recent years, inspired by the success of physics-informed neural networks (PINNs) in machine learning and scientific computing~\cite{sirignano2018DGM,raissi2019pinns}, PINNs-based methods 
are increasingly being used for reachability analysis.
However, PINNs are not a silver bullet; direct application of PINNs
to HJI-VI PDEs does not necessarily yield high-fidelity value functions~\cite{sharpless2026linearsupervision,teoh2025mpcadversarial,feng2025mpcreachability},
a trend that is also present in other domains such as fluid dynamics~\cite{chuang2023predictive}. 
In order to overcome these challenges, recent 
state-of-the-art (SoTA) PINNs-based reachability solvers~\cite{sharpless2026linearsupervision,teoh2025mpcadversarial,feng2025mpcreachability,bansal2020deepreach,singh2025exact} 
incorporate both architectural improvements
(e.g., periodic activation functions~\cite{sitzmann2020periodic}
and exact boundary constraints~\cite{singh2025exact}), 
in addition to algorithmic techniques, including time-based curriculum learning~\cite{bansal2020deepreach,krishnapriyan2021characterizing}, 
linear semi-supervision~\cite{sharpless2026linearsupervision},
and MPC-guided sampling with multi-stage training pipelines~\cite{teoh2025mpcadversarial,feng2025mpcreachability}. 
The result is a collection of complex algorithms that are non-trivial to implement and challenging to tune.

\begin{figure}[t]
    \centering
    \begin{subfigure}{.49\linewidth}
        \includegraphics[width=\linewidth]{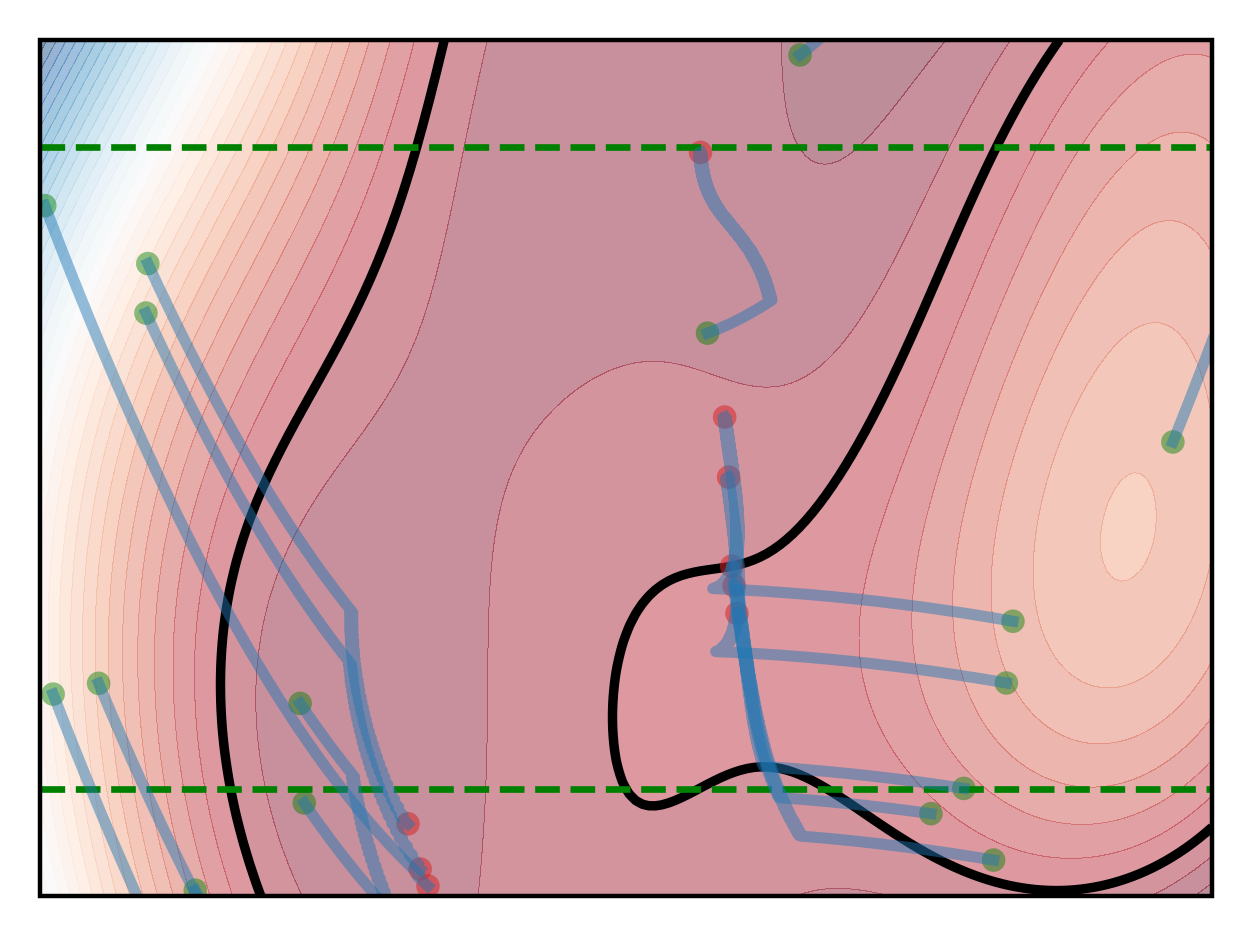}
        \caption{Trajectories at initialization.}
    \end{subfigure}
    \begin{subfigure}{.49\linewidth}
        \includegraphics[width=\linewidth]{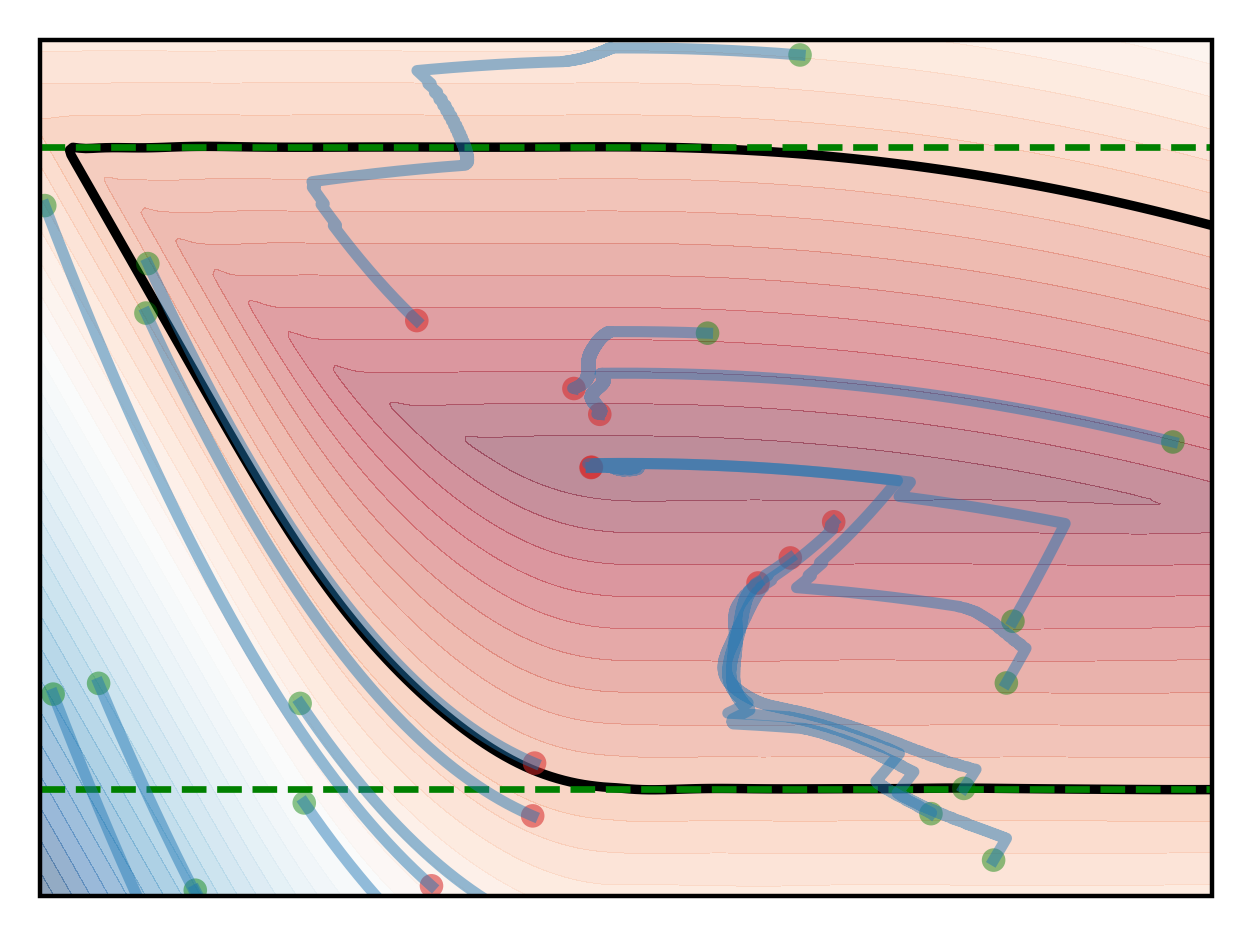}
        \caption{Trajectories at convergence.}
    \end{subfigure}
    \caption{A plot of sample trajectories from the 2D vertical drone problem~\cite{feng2025mpcreachability}. 
    The green dotted lines represent the floor and ceiling that the system should avoid, the black line represents the learned 
    safety value function zero-level contour, and the blue lines illustrate sampled trajectories. For each sample trajectory, the green and red dots represent the start and end points, respectively.
    }
    \label{fig:trajectory_illustration}
\end{figure}

A central theme that emerges from the existing literature is that the
sampling distribution over collocation points plays a crucial role in the performance of PINNs-based solvers;
moving away from uniform grid-based distributions towards 
problem-adaptive sampling is critical. 
However, designing a suitable adaptive sampling scheme can introduce substantial algorithmic complexity in PINNs methods.
This motivates the central question of our work:
\begin{center}
\itshape
\textbf{Q:} Can one design a \textbf{simple} adaptive sampling strategy
for PINNs-based HJ reachability analysis
that achieves near SoTA performance on high-dimensional problems?
\end{center}
We provide an affirmative answer via \MethodName{} (\MethodAbbrv{}),
drawing inspiration from backward stochastic differential equation (SDE) methods for solving PDEs~\cite{han2018bsde,raissi2024forward,nusken2023interpolating,park2025integration}. 
\MethodAbbrv{} constructs an adaptive sampling distribution (cf.~\Cref{fig:trajectory_illustration}) by (i) using the current value function to compute control and disturbance signals that guide trajectory rollouts, and (ii) injecting stochasticity around these trajectories by modeling the system as an SDE with a tunable noise parameter. 
We show that optimizing the PINNs loss under this sampling distribution is sufficient to achieve 
competitive performance
across a range of HJ reachability benchmarks. 
Specifically, compared with the SoTA MPC-guided solver~\cite{feng2025mpcreachability},
\MethodAbbrv{} often matches and can even improve performance on safety metrics, while consistently achieving lower relative $L_2$ error.
Moreover, \MethodAbbrv{} is straightforward to implement, introduces minimal overhead beyond PINNs, 
and avoids the need for multi-stage training pipelines, curriculum learning, and auxiliary supervision from MPC or linear approximations.

Our paper is organized as follows. 
The problem formulation and HJ reachability framework is introduced in
\Cref{sec:problem_formulation},
with \Cref{sec:background} containing detailed discussion on 
existing PINNs-based HJ reachability solvers.
\Cref{sec:steer2reach} introduces our \MethodAbbrv{}
approach, \Cref{sec:experiments} provides a detailed experimental evaluation,
and \Cref{sec:conclusion} concludes.

\section{Problem Formulation}
\label{sec:problem_formulation}

Consider a dynamical system with state $x\in \R^n$, control $u \in \calU$, and disturbance $d \in \calD$, where the state evolves as:
\begin{equation}
    \dot x=f(x,u,d),\quad u\in\calU,\; d\in \calD. \label{eq:dynamics}
\end{equation}
In this work, we aim to compute the \emph{Backward Reachable Tube} (BRT), 
denoted $\calB$, which contains the initial states for which 
the system \eqref{eq:dynamics} will inevitably enter a given failure set $\calL$ under some worst-case disturbance within a time horizon $\calI=[0,T]$,
no matter what control actions are taken.
Following standard convention, the BRT is defined as:
\begin{equation}\label{eq:brt}
    \calB = \{x \mid \forall u \in \sfU,\exists d \in \sfD,\tau\in\calI \,\,\text{s.t.}\,\, \Phi^{u,d}_{x,t}(\tau)\in \calL \},
\end{equation}
where $\Phi^{u,d}_{x,t}(\tau)$ represents the state at time $\tau$ given initial state $x(t)=x$ at time $t \in \calI$ and applying the control $u(\cdot) \in \sfU$ and disturbance $d(\cdot) \in \sfD$ over $[t, \tau]$. 
Here, the set $\sfU$ (resp.~$\sfD$) denotes the space
of measurable functions mapping $\calI \mapsto \calU$
(resp.~$\calI \mapsto \calD$).
We assume that the failure set $\calL=\set{x \in \R^n \mid \ell(x)\leq 0}$
is described as the zero sub-level set of a Lipschitz-continuous function $\ell(x):\R^n \mapsto \R$,
which we typically take as a signed-distance function.

In avoid problems, where the set $\calL$ is composed of unsafe states, the BRT contains all states from which the system will inevitably enter the failure set despite best effort controls under worst case disturbance, and hence should be avoided. 
Conversely, in reach problems
where the set $\calL$ denotes the target set,
we reverse the role of the disturbance and control, and consequently the BRT contains all states from which the system, acting optimally, will eventually reach the target set despite worse case disturbances. In addition to the BRT, we aim to find an optimal safe control policy, $u^*(\cdot)$ for states outside the BRT, which ensures the system avoids the target set despite worst case disturbances.

\subsection{Hamilton-Jacobi-Isaacs Variational Inequality}

The standard method for computing the BRT arises from Hamilton-Jacobi (HJ) reachability~\cite{mitchell2005reachability,bansal2017reachability}, which we
summarized the key ideas here.
HJ reachability formulates $\calB$ as the zero sub-level set of the value function,
\begin{equation}\label{eq:value}
    V(x,t) = \max_{u \in \sfU} \min_{d \in \sfD} J(x,t,u,d),
\end{equation}
where $J(x,t,u,d) = \min_{\tau \in [t, T]} \ell\left(\Phi_{x,t}^{u,d}(\tau)\right)$ measures the minimum distance to the set $\calL$, measured via $\ell$, along a trajectory $(\Phi_{x,t}^{u,d}(\tau))_{\tau=t}^{T}$. 
The value function $V(x,t)$ can be computed as the viscosity solution to the Hamilton-Jacobi-isaacs Variational Inequality (HJI-VI),
which is described by two key equations:
\begin{align}
    \!R_V(x, t) &= \min\{ \partial_t V(x,t) + H(x,t), \ell(x)-V(x,t) \}, \label{eq:residual} \\
    H(x, t) &= \max_{u \in \calU} \min_{d \in \calD} \ip{\nabla_xV(x,t)}{f(x,u,d)}. \label{eq:hji_h}
\end{align}
From HJI-VI, the value function $V(x, t)$ satisfies:
\begin{align}
    R_V(x, t) = 0, \quad V(x, T) = \ell(x), \quad x \in \R^n, t \in \calI. \label{eq:HJI_VI}
\end{align}

Given a value function $V(x,t)$ satisfying \eqref{eq:HJI_VI}, we compute the BRT, optimal control $u^*$, and optimal disturbance $d^*$, as:
\begin{align}
    \calB &=\Set{x \in \R^n \mid V(x,0)\leq 0}, \nonumber \\
    u^*(x,t)&=\argmax_{u \in \calU }\min_{d \in \calD} \ip{\nabla V(x,t)}{f(x,u,d)}, \label{eq:optimal_control} \\
    d^*(x,t; u)&=\argmin_{d \in \calD} \ip{\nabla V(x,t)}{f(x,u(x, t),d)}. \label{eq:optimal_disturbance}
\end{align}

\section{Background and Related Work: PINNs-based Hamilton-Jacobi Reachability Solvers}
\label{sec:background}

An exact solution to an HJI-VI equation is not computationally tractable for systems where the number of states exceeds a modest dimension. 
Consequently, recently machine learning methods, specifically 
physics-informed neural networks (PINNs)~\cite{raissi2019pinns,bansal2020deepreach}, 
have been used to learn approximate value functions. In this section, we outline 
these approaches, as they form the basis for our method.

Specifically, we consider learning approximate solutions to the HJI-VI equation \eqref{eq:HJI_VI} 
over a compact domain $\Omega \times \calI$, with $\Omega\subseteq \R^n$, by parameterizing the value function $V_\theta(x,t)$ as an element of a
rich function class $\calV=\set{V_\theta(x,t)|\theta\in \Theta}$ (e.g., $\theta \in \Theta$ represents the weights of a neural network).
This parameterization induces two natural loss functions over the parameter space $\Theta$:
\begin{align}
    \calL_{\text{VI}}(\theta; \mu) &= \underset{(x, t) \sim \mu}{\E}\left[ R_{V_\theta}(x,t)^2\right], \label{eq:loss_VI} \\
    \calL_{\text{bnd}}(\theta; \mu') &= \underset{x \sim \mu'}{\E}[(V_\theta(x,T)-\ell(x))^2]. \label{eq:loss_boundary}
\end{align}
where $\mu$ and $\mu'$ are measures over $\Omega \times \calI$ and $\Omega$, respectively. 
We remark here that while we utilize the
square-loss in $\calL_{\text{VI}}$ and
$\calL_{\text{bnd}}$, other losses such as absolute value~\cite{bansal2020deepreach}
or a robust adversarial loss~\cite{wang2022infinitylosspinns}
can also be used.

\subsection{Basic PINNs Loss and the DeepReach Solver}
\label{sec:background:deepreach}
With both \eqref{eq:loss_VI} and \eqref{eq:loss_boundary} in place, we can now describe
the basic PINNs loss:
\begin{align}\label{eq:loss_pinns}
    \calL_{\text{PINNs}}(\theta;\lambda)= \calL_{\text{VI}}(\theta; U(\Omega \times \calI)) + \lambda \calL_{\text{bnd}}(\theta; U(\Omega)),
\end{align}
where $U(\calS)$ is the uniform measure over the set $\calS$.

DeepReach~\cite{bansal2020deepreach} is an approach which builds on this PINNs loss \eqref{eq:loss_pinns}, but proposes several modifications:
(i) \emph{time-based curriculum:}
the network $V_\theta$ is first initialized 
minimizing the loss $\calL_{\text{bnd}}$, followed by
optimizing the PINNs loss but with a 
curriculum of samples, i.e., using $U(\Omega \times [t_i, T])$ in the $\calL_{\text{VI}}$ loss, where $t_i$ decays to $0$ as the training progresses,
(ii) \emph{sinusoidal networks:} MLPs with sinusoidal activation functions~\cite{sitzmann2020periodic} 
are used to parameterize $V_\theta$.
Recent follow up work \cite{singh2025exact}
removes the $\calL_{\text{bnd}}$ loss, and
instead parameterizes
$V_\theta(x, t) = \ell(x) - (T-t) \phi_\theta(x, t)$, where $\phi_\theta(x, t)$ is an unconstrained network.
This satisfies $V_\theta(x, T) = \ell(x)$ by construction, and allows one to remove $\calL_{\text{bnd}}$.

\subsection{Residual Adaptive Distribution (RAD)}\label{sec:methods:rad}

One of the issues with the standard PINNs loss \eqref{eq:loss_pinns} is that the sampling of collocation points $(x, t)$ is uniform
over the domain $\Omega \times \calI$ and non-adaptive.
There is a large body of literature in the
PINNs community on adaptive sampling schemes
to overcome this issue~\cite{nabian2021importancesampling,wu2023comprehensiveadaptive}. 
Here, we describe one canonical sampling method,
as it will serve as an additional baseline.
The residual-based adaptive distribution (RAD) method~\cite{wu2023comprehensiveadaptive} focuses collocation sampling in areas with high PDE residuals by sampling points according to a probability density function $\mu_\theta^{\text{RAD}}(x, t)$ proportional to the PDE residual. 
Defining $\e_\theta(x, t) = \abs{R_{V_\theta}(x, t)}$, we have $\mu_\theta^{\text{RAD}}(x, t)\propto \frac{\varepsilon_\theta^k(x, t)}{\E_{(x,t) \sim \mu_0}[\varepsilon_\theta^k(x, t)]}+c,$
where $k, c \geq 0$ are two hyperparameters that
affect the shape of $\mu_\theta^{\text{RAD}}$, 
and $\mu_0$ is a base measure (e.g., uniform).
We utilize the Monte-Carlo method of generating samples
described in \cite[Sec.\ 2.3.2]{wu2023comprehensiveadaptive}
(also cf.\ \Cref{sec:appendix:radalgo}).

\subsection{MPC-Guided DeepReach Solver}\label{sec:background:mpcdeepreach}

A recent state-of-the-art solver, MPC-guided DeepReach~\cite{feng2025mpcreachability} (MPC-DeepReach),
improves upon the original DeepReach work by
leveraging MPC-guided sampling and supervised learning.
Specifically, MPC-DeepReach utilizes a sampling-based MPC to generate a dataset of approximate state/value pairs,
which is then used in a three step optimization process involving:
(i) a supervised pre-training phase, 
(ii) a joint curriculum training phase, and
(iii) a fine-tuning phase to reduce false positives. 
In the supervised pre-training phase, the approximate value function generated via MPC is used to warm-start the neural network. 
Then, in the following curriculum training and fine-tuning phases, 
the PDE residual and MPC regression losses are jointly minimized using an adaptive weighting function,
with the MPC-generated dataset periodically rebuilt from the most
current value function. 
The fine-tuning phase utilizes an imbalanced regression loss that amplifies the false positive residuals to help reduce false positive predictions in the learned value function.
MPC-DeepReach was later extended to two-player
zero-sum games in \cite{teoh2025mpcadversarial}.
We note that this technique of computing an approximate $V(x, t)$ to semi-supervise
learning the safety value function is also explored in \cite{sharpless2026linearsupervision}, 
where the Hopf formula is used to efficiently compute $V(x, t)$ under Jacobian linearized dynamics.
In addition to MPC-guided semi-supervision, MPC-DeepReach 
also uses the methods described in~\Cref{sec:background:deepreach},
plus a gradient normalization method for loss weight balancing.

\section{\MethodName{}: Forward Trajectory Steering for Reachability Analysis}
\label{sec:steer2reach}

While \Cref{sec:background} describes a variety of PINNs-based approaches to solving the HJI-VI problem,
we see one crucial commonality shared among these approaches. In particular, whether through
curriculum learning in DeepReach,
adaptive sampling for RAD, or 
through MPC solutions in MPC-DeepReach,
the core principle is to modify the sampling distribution over the collocation
points away from the uniform distribution on $\Omega \times \calI$.
While this literature shows that non-uniform sampling
is crucial for learning accurate value functions,
state-of-the-art approaches for sampling are becoming increasingly complex to implement, involving
several training phases that each need to be separately tuned for the best performance.
In this section, we propose a natural non-uniform sampling approach---which we coin \MethodName{}---that is
very simple to implement, while largely retaining the performance of existing approaches.

Our approach is heavily inspired by the rich literature on 
backward stochastic differential equations (BSDE)-based techniques for solving PDEs~\cite{han2018bsde,raissi2024forward,nusken2023interpolating,park2025integration}.
The idea behind BSDE-based solvers is to first construct a pair of forward/backward SDEs derived from the target PDE,
such that the value of the backward SDE encodes the PDE solution at a desired evaluation point.
For HJI-VI PDEs, which are more complex than the standard semilinear parabolic PDEs that are typically
considered in the BSDE literature, this is done via \emph{reflected} BSDEs~\cite{el1997reflected,buckdahn2011stochastic}.

\subsection{Reflected BSDEs for HJI-VI PDEs}
To set the stage, we first start by extending
the deterministic dynamics~\eqref{eq:dynamics} to
an SDE driven by $\sigma : \R^n \times \calU \times \calD \mapsto \R^{n \times q}$. Given an admissible $(u, d) \in \sfU \times \sfD$,\footnote{
$\sfU, \sfD$ here are overloaded as the set of signals $u(t), d(t)$ that are progressively measurable w.r.t.\ $(\calF_t)_{t \geq 0}$, where $\calF_t = \sigma( (W_s)_{s=0}^{t} )$.} 
for $\tau \in [t, T]$:
\begin{multline*}
    \rmd X^{t,x;u,d}_{\tau} = f(X^{t,x;u,d}_{\tau}, u_\tau, d_\tau) \rmd \tau \\+ \sigma(X^{t,x;u,d}_{\tau}, u_\tau, d_\tau) \rmd W_\tau,
\end{multline*}
where $X_t^{t,x;u,d} = x$ and $(W_t)_{t \geq 0}$ denotes standard
Brownian motion in $\R^q$.
We then define the modified stochastic
value function $V^\sigma(x, t)$ as:
$$
    V^\sigma(x, t) = \max_{u \in \sfU} \min_{d \in \sfD} \min_{\tau \in \calT_t} \E[ \ell(X^{t,x;u,d}_\tau) ],
$$
where $\calT_t$ denotes the set of all stopping times contained in the interval
$[t, T]$.
We note that under sufficient regularity we can interpret $V^\sigma \to V$ as $\sigma \to 0$.
The results of \cite{el1997reflected,buckdahn2011stochastic}
combined state that for the forward process 
$(X_\tau^{t,x;u,d})_{\tau=t}^{T}$
and the reflected upper obstacle BSDE $(Y_\tau^{t,x;u,d}, Z_\tau^{t,x;u,d}, K_\tau^{t,x;u,d})_{\tau=t}^{T}$:
\begin{align*}
Y_\tau^{t,x;u,d}
&= \ell\!\left(X_T^{t,x;u,d}\right)
-\int_\tau^T Z_s^{t,x;u,d}\, \rmd W_s\\
&\hspace{10em}- \bigl(K_T^{t,x;u,d} - K_\tau^{t,x;u,d}\bigr), \\
0 &= \int_t^T
\bigl(
\ell(X_s^{t,x;u,d}) - Y_s^{t,x;u,d}
\bigr)\, \rmd K_s^{t,x;u,d},
\end{align*}
with $\tau \in [t, T]$, $Y_\tau^{t,x;u,d}
\le \ell\!\left(X_\tau^{t,x;u,d}\right)$,
$K_t^{t,x;u,d} = 0$ and $(K_\tau^{t,x;u,d})_{\tau=t}^{T}$ non-decreasing, we have the identity
$V^\sigma(x, t) = \max_{u \in \sfU} \min_{d \in \sfD} Y_{t}^{t,x;u,d}$.

Hence, the value function $V^\sigma$ can be recovered numerically
by finding a solution to the reflected BSDE. 
However, while various methods have been devised to solve reflected
BSDEs~\cite{gobet2008numerical,bender2017primal,bayraktar2024deepsignature}, the reflected conditions arising from the variational inequality leads to complex algorithms. Hence, instead of directly
trying to solve the reflected BSDE,
we take inspiration from recent work~\cite{park2025integration}
which shows that utilizing the forward SDE
$(X_t^{0,x;u,d})_{t=0}^{T}$ to steer PINNs optimization 
via collocation point sampling
yields competitive performance with
SoTA Heun-based BSDE solvers,
and is very simple to implement on top of 
existing PINNs solvers.

\subsection{The \MethodName{} Loss Function and Optimization}
\label{sec:steer2reach}

We now present our main \MethodName{} approach. In order to use the forward
SDE as a sampling distribution, we need to 
choose an admissible pair $(u, d) \in \sfU \times \sfD$. To do this, we consider a family of pairs $(u_\theta, d_\theta)$ 
parameterized by the value function $V_\theta(x, t)$. Specifically, we set
$(u_\theta, d_\theta)$ to be the optimal
control and disturbance assuming that $V = V_\theta$. From \eqref{eq:optimal_control} and
\eqref{eq:optimal_disturbance}, 
\begin{align}
    u_\theta(x,t)&=\argmax_{u \in \calU }\min_{d \in \calD} \ip{\nabla V_\theta(x,t)}{f(x,u,d)}, \label{eq:control_theta} \\
    d_\theta(x,t)&=\argmin_{d \in \calD} \ip{\nabla V_\theta(x,t)}{f(x,u_\theta(x, t),d)}. \label{eq:disturbance_theta}
\end{align}
We then define $X^{x;\theta}_t \equiv X_t^{0,x;u_\theta,d_\theta}$.
We use this forward SDE to induce measures
$\mu_\theta$ (on $\R^n \times \calI$)
and $\mu'_\theta$ (on $\R^n$)
as $\mu_\theta = \mathrm{Law}(( X_t^{x;\theta}, t ))$ and $\mu_\theta' = \mathrm{Law}( X_T^{x;\theta} )$, where $t \sim U(\calI)$,
$x \sim \mu_0$, and $x \perp t$.
Here, $\mu_0$ is a fixed distribution over
initial conditions.
These measures then give rise to the
\MethodName{} loss:
\begin{align}
    \calL_{\text{S2R}}(\theta; \theta_{\text{samp}}, \lambda) = \calL_{\text{VI}}(\theta; \mu_{\theta_{\text{samp}}}) + \lambda \calL_{\text{bnd}}(\theta; \mu'_{\theta_{\text{samp}}}). \label{eq:S2R_loss}
\end{align}

The \MethodName{} loss is optimized by using 
the current value function parameters $\theta_t$
as the sampling parameters. Specifically, at optimization iteration $t$, we define the function $\calL_{\text{S2R},t}(\theta) \equiv \calL_{\text{S2R}}(\theta; \theta_t, \lambda)$, and
utilize gradients $\nabla \calL_{\text{S2R},t}(\theta_t)$ to update the parameters to $\theta_{t+1}$:
\begin{align}
    \theta_{t+1} = \mathrm{UpdateRule}( \theta_t, \nabla \calL_{\text{S2R},t}(\theta) ), \label{eq:repeated_GD}
\end{align}
where $\mathrm{UpdateRule}$ is any gradient-based optimization rule such as gradient descent, L-BFGS, Adam, etc.
This family of optimization problems, where the underlying distribution changes as a function of the current parameter value $\theta_t$, is well-studied in the
machine learning literature under the umbrella of
performative prediction~\cite{perdomo2020performative}.
In this literature, the type of update scheme in \eqref{eq:repeated_GD} is known as \emph{repeated gradient descent (RGD)}. 

An alternative to RGD \eqref{eq:repeated_GD} is to directly minimize the loss
$\calL_{\text{S2R,full}}(\theta) \equiv \calL_{\text{S2R}}(\theta; \theta, \lambda)$.
In applications where the control and disturbance
sets are e.g., box constraints, the gradients of $u_\theta, d_\theta$ are zero almost everywhere, and hence in practice
using gradient methods to minimize $\calL_{\text{S2R,full}}(\theta)$
is functionally equivalent to \eqref{eq:repeated_GD}.
We further experimented with using direct minimization by adding extra regularization terms on
$u, d$ 
to \eqref{eq:control_theta}, \eqref{eq:disturbance_theta}
that ensure the $u_\theta, d_\theta$ would have non-zero
gradients that could be used to guide optimization
in the early stages. However, we found this direct minimization with regularization to be quite unstable to train; we leave effectively leveraging smoothed gradients w.r.t.\ the measures $\mu_{\theta}, \mu_{\theta}'$ to future work.

We summarize the full \MethodName{} algorithm in~\Cref{alg:fspinns}.
We utilize the standard Euler-Maruyama (EM) integrator
to rollout the forward SDE $X_t^{x;\theta}$.
We implement RGD updates by using a stop-gradient operator 
on the forward SDE trajectories.
On top of the vanilla PINNs approach \eqref{eq:loss_pinns},
\MethodAbbrv{} introduces two new hyperparameters surrounding
the forward SDE: (a) the noise injection $\sigma$ (we utilize the constant function $\sigma(x, u, d) \equiv \sigma \cdot I_n$),
and (b) the integration timestep $\Delta t$ for EM 
integration.
We study the sensitivity of \MethodAbbrv{} to 
both the noise injection parameter $\sigma$ 
and the integration time $\Delta t$ in \Cref{sec:experiments:ablation}.

\begin{algorithm}[htb]
    \caption{The \MethodName{} (\MethodAbbrv{}) Algorithm}
    \label{alg:fspinns}
    \begin{algorithmic}[1]
        \small
        \Input Initial $\theta$, 
        boundary loss weight $\lambda$, 
        optimization iterations $N_{\text{iter}}$, 
        forward SDE noise $\sigma$,
        forward SDE discretization $\Delta t$, 
        number of batched rollouts $B_{\text{traj}}$, 
        loss batch size $B_{\text{pde}}$.
        \Output Value function parameters $\theta$.
        \State Set $N_{\text{traj}} = T/\Delta t$. \algcomment{ // Assume evenly divides}
        \For{$j=0, \dots, N_{\text{iter}}-1$}
            \State \algcomment{/* Batched forward SDE rollout */}
            \State Sample $x_i[0] \stackrel{\text{i.i.d.}}{\sim} U(\Omega),\quad i=1,\dots,B_{\text{traj}}$.
            \State Sample $\xi_i[0:N_{\text{traj}}-1] \stackrel{\text{i.i.d.}}{\sim} \mathsf{N}(0,I_n)$.
            \State Set $t_i = (\Delta t \cdot k)_{k=0}^{N_{\text{traj}}}$.
            \State Define $F_\theta(x, t) = f(x, u_\theta(x, t), d_\theta(x, t))$.
            \State \algcomment{/* Euler-Maruyama integration */}
            \For{$k=0,\dots,N_{\text{traj}}-1$}
                \State \algcomment{/* Implement with e.g., jax.vmap */}
                \State $x_i[k+1]= x_i[k]+ \Delta t \cdot F_\theta(x_i[k], t_i[k]) + \sqrt{\Delta t} \sigma \xi_i[k]$.
            \EndFor
            \State $x_i =\text{stop\_grad}(x_i)$. \texttt{ // See Equation \eqref{eq:repeated_GD}}

            \State \algcomment{/* PINNs loss computation */}
            \State Set $x_{\text{all}} = \text{hstack}([x_1, \dots, x_{B_{\text{traj}}}])$.
            \State Set $t_{\text{all}} = \text{hstack}([t_1, \dots, t_{B_{\text{traj}}}])$.
            \State Set $(x_b,t_b)=\text{rand\_permute}(\text{vstack}([x_{\text{all}}, t_{\text{all}}]))[0:B_{\text{pde}}]$.
            \State Set $x_{\text{bnd}} = \text{rand\_permute}([x_1[T], \dots, x_{B_{\text{traj}}}[T]])[0: B_{\text{pde}}]$.
            \State Define $\calL_{\text{VI}}(\theta) =\sum_{m=0}^{B_{\text{pde}}-1} R_{V_\theta}^2(x_b[m],t_b[m])$.
            \State Define $\calL_{\text{bnd}}(\theta) =  \sum_{m=0}^{B_{\text{pde}}-1} ( \ell(x_{\text{bnd}}[m]) - V_\theta(x_{\text{bnd}}[m], T) )^2$.
            \State Define $\calL_{\text{S2R},j}(\theta) = \calL_{\text{VI}}(\theta_j) + \lambda\calL_{\text{bnd}}(\theta_j)$.
            \State $\theta_{j+1} \leftarrow \mathrm{UpdateRule}(\theta_j,  \nabla_\theta \calL_{\text{S2R},j}(\theta_j))$.
        \EndFor
        \State \textbf{Return:} Parameters $\theta_{N_{\text{iter}}}$.
    \end{algorithmic}
\end{algorithm}

\subsection{Value Function Consistency Constraints}\label{sec:s2rhc}

Finally, we describe our last technique, which draws inspiration from the work of \cite{singh2025exact,choi2023forward}. 
By the definition of $V(x, t)$ in \eqref{eq:value}, it is immediate to verify that
$V(x, t) \leq \ell(x)$ holds for all $x \in \R^n$.
At the same time, we also have the terminal constraint
$V(x, T) = \ell(x)$ 
from \eqref{eq:HJI_VI}.
Hence, we propose to \emph{simultaneously}
encode these two constraints in our parameterization
via:
\begin{align}\label{eq:hcvalue}
    V_\theta(x,t)= \ell(x)-(T-t)\rho(\phi_\theta(x,t)),
\end{align}
where $\rho:\R\mapsto \R_{\geq 0}$ is a fixed (non-learnable) function with non-negative outputs (e.g.\ a quadratic
$\rho(x) = x^2$), and $\phi_\theta(x, t)$ is an
unconstrained network.
Note that this parameterization, in addition to removing the
need for the boundary loss $\calL_{\text{bnd}}$ in \eqref{eq:S2R_loss}
and \Cref{alg:fspinns}, also
enforces value function consistency 
$V_\theta(x, t) \leq \ell(x)$ for all $x \in \R^n$,
excluding the possibility of representing solutions
$V_\theta$ where the value \emph{grows} as $t$ moves away from the final time $T$.

\section{Experiments}
\label{sec:experiments}

\begin{figure*}[!tb]
    \centering
    \begin{subfigure}[h]{\linewidth}
        \caption{Comparison plots of the 
        learned safety value function $V_\theta(x, 0)$ for the 2D vertical drone problem.}\label{fig:vd}
        \includegraphics[width=\linewidth]{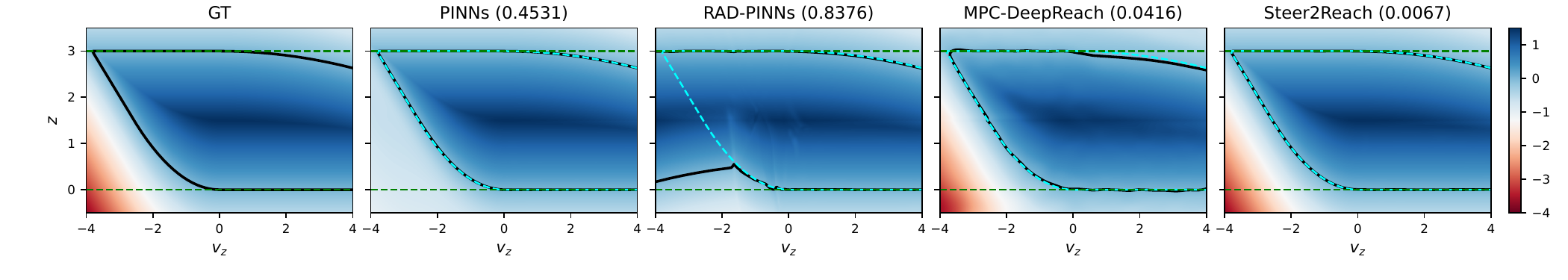}
    \end{subfigure}
    \hfill
    \begin{subfigure}[b]{\linewidth}
        \caption{Comparison plots of the
        learned safety value function $V_\theta(x, 0)$ for the 40D publisher-subscriber problem 
        evaluated on the slice $x=(x_0, x_i, \dots, x_i)$.
        }
        \includegraphics[width=1\linewidth]{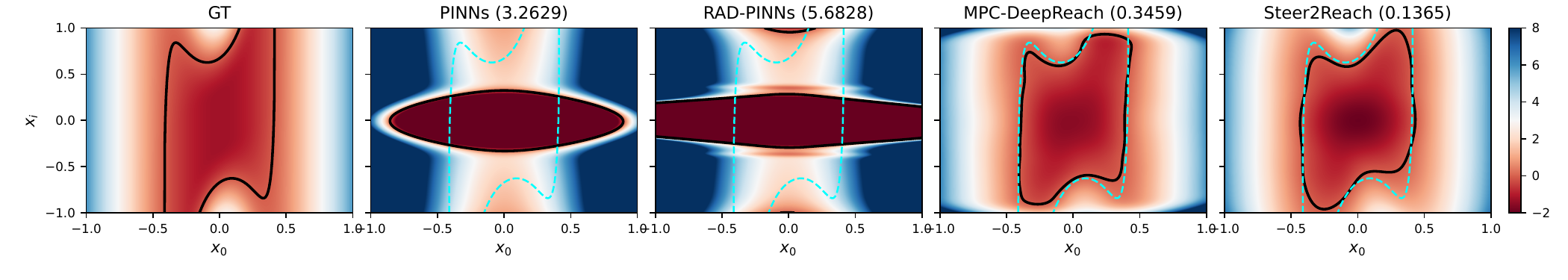}
    \end{subfigure}
    \caption{Comparison plots of the learned safety value function $V_\theta(x, t)$ at $t=0$ with the zero level set outlined as a black line. The green dotted line in~(a) represent the floor and ceiling that the system should avoid and the white dotted line represents the ground truth zero level contour. We include the RL2 error of each method in the title for reference.}
    \label{fig:valueplots}
\end{figure*}

In this section, we compare \MethodName{} against standard PINNs, 
PINNs with RAD sampling~\cite{wu2023comprehensiveadaptive},
and MPC-DeepReach~\cite{feng2025mpcreachability}.
Specifically:
\begin{enumerate}[(\alph*)]
    \item \textbf{PINNs:} The PINNs loss from~\eqref{eq:loss_pinns} is minimized with the hard-constrained network described in~\Cref{sec:s2rhc}.
    This method is most similar to DeepReach with exact boundary constraints~\cite{singh2025exact},
    although we do not label it as such since we do not utilize
    DeepReach's time-based curriculum and sinusoidal networks.
    \item \textbf{RAD-PINNs:} We minimize the PINNs loss \eqref{eq:loss_pinns}, utilizing the 
    RAD~\cite{wu2023comprehensiveadaptive} measure $\mu_\theta^{\text{RAD}}$ described in \Cref{sec:methods:rad}, with hyperparameters $k=c=1$.
    \item \textbf{MPC-DeepReach:} We use the MPC-DeepReach (MPC-DR) codebase~\cite{deepreach2024} to learn safety value functions
    with the method described in \Cref{sec:background:mpcdeepreach}.
    \item \textbf{\MethodName:} Our approach as described in 
    \Cref{sec:steer2reach} and \Cref{alg:fspinns}.
\end{enumerate}

We evaluate these methods on five reachability benchmarks described in~\cite{bansal2020deepreach,feng2025mpcreachability}: 2D Vertical Drone, 3D Pursuit-Evade, 7D F1Tenth, 13D Quadrotor, and 40D Publisher-Subscriber. 
The dynamics and unsafe sets $\calL$ for each problem are detailed in \Cref{appendix:sec:problems}.

\begin{table*}[tb]
    \centering

\begin{tabular}{ll|ccccc}
    \toprule
    Problem & Model & Precision & IOU & TV & RL2@$t{=}0$& Runtime[m]  \\
    \midrule
    \multirow{4}{*}{2D VerticalDrone} & PINNs & $0.8328 \pm 0.2799$ & $0.8266 \pm 0.2787$ & $50.8891 \pm 12.8575$ & $0.3545 \pm 0.4371$ & 0.41 \\
     & RAD-PINNs & $0.8590 \pm 0.2215$ & $0.8506 \pm 0.2187$ & $52.9285 \pm 8.8048$ & $0.4277 \pm 0.3767$ & 0.61 \\
     & MPC-DR & $\mathbf{0.9833 \pm 0.0015}$ & $\underline{0.9688 \pm 0.0039}$ & $\mathbf{57.5331 \pm 0.1758}$ & $\underline{0.1020 \pm 0.0290}$ & 1.34 \\
     & S2R & $\underline{0.9780 \pm 0.0039}$ & $\mathbf{0.9745 \pm 0.0037}$ & $\underline{57.4632 \pm 0.2630}$ & $\mathbf{0.0497 \pm 0.0014}$ & 1.04 \\
    \midrule
    \multirow{4}{*}{PursuitEvade 3D} & PINNs & $0.9999 \pm 0.0001$ & $0.9702 \pm 0.0053$ & $91.7998 \pm 0.0579$ & $0.0617 \pm 0.0071$ & 1.11 \\
     & RAD-PINNs & $\mathbf{1.0000 \pm 0.0000}$ & $0.9587 \pm 0.0104$ & $\mathbf{92.3046 \pm 0.0700}$ & $0.0699 \pm 0.0168$ & 1.90 \\
     & MPC-DR & $0.9995 \pm 0.0000$ & $\mathbf{0.9910 \pm 0.0003}$ & $91.9236 \pm 0.0060$ & $\underline{0.0360 \pm 0.0087}$ & 6.30 \\
     & S2R & $\mathbf{1.0000 \pm 0.0000}$ & $\underline{0.9854 \pm 0.0005}$ & $\underline{91.9665 \pm 0.0082}$ & $\mathbf{0.0271 \pm 0.0019}$ & 3.24 \\
    \midrule
\multirow{4}{*}{F1Tenth 7D} & PINNs & $0.7058 \pm 0.0742$ & $0.7012 \pm 0.0727$ & $66.8001 \pm 6.7809$ & - & 46.60 \\
 & RAD-PINNs & $0.4211 \pm 0.0382$ & $0.4166 \pm 0.0378$ & $40.1880 \pm 3.7511$ & - & 52.65 \\
 & MPC-DR & $\mathbf{0.9794 \pm 0.0011}$ & $\mathbf{0.9603 \pm 0.0031}$ & $\mathbf{82.0896 \pm 0.1341}$ & - & 306.27 \\
 & S2R & $\underline{0.7500 \pm 0.0220}$ & $\underline{0.7364 \pm 0.0200}$ & $\underline{71.8773 \pm 2.0967}$ & - & 105.78 \\
    \midrule
    \multirow{4}{*}{Quadrotor 13D} & PINNs & $0.9141 \pm 0.0071$ & $0.9098 \pm 0.0066$ & $87.2352 \pm 0.6200$ & - & 47.69 \\
     & RAD-PINNs & $0.8997 \pm 0.0018$ & $0.8955 \pm 0.0019$ & $86.1629 \pm 0.0674$ & - & 52.98 \\
     & MPC-DR & $\mathbf{0.9896 \pm 0.0001}$ & $\mathbf{0.9884 \pm 0.0001}$ & $\mathbf{94.6099 \pm 0.0031}$ & - & 118.46 \\
     & S2R & $\underline{0.9773 \pm 0.0191}$ & $\underline{0.9535 \pm 0.0113}$ & $\underline{92.3880 \pm 1.6725}$ & - & 87.33 \\
    \midrule
    \multirow{4}{*}{PublisherSubscriber 40D} & PINNs & $0.0000 \pm 0.0000$ & $0.0000 \pm 0.0000$ & $38.8330 \pm 0.4232$ & $1.7727 \pm 0.0756$ & 18.05 \\
     & RAD-PINNs & $0.0000 \pm 0.0000$ & $0.0000 \pm 0.0000$ & $23.1759 \pm 9.1214$ & $2.2479 \pm 0.0216$ & 23.22 \\
     & MPC-DR & $\underline{0.9971 \pm 0.0007}$ & $\underline{0.9831 \pm 0.0013}$ & $\underline{39.7366 \pm 0.0018}$ & $\underline{0.5676 \pm 0.1539}$ & 82.85 \\
     & S2R & $\mathbf{0.9979 \pm 0.0000}$ & $\mathbf{0.9878 \pm 0.0004}$ & $\mathbf{39.9131 \pm 0.0005}$ & $\mathbf{0.0983 \pm 0.0266}$ & 50.63 \\
    \bottomrule
\end{tabular}
    \caption{A table of evaluation metrics averaged over three random initializations. Settings with no ground truth are denoted with - . In addition, we mark the highest performance for each problem/metric in \textbf{bold} and the runner up with an \underline{underline}.
    In all cases, \MethodName{} utilizes  $\sigma = .01$ and a $\Delta_t$ such that $N_{\text{traj}} = 50$ (cf.~\Cref{alg:fspinns}).
    } 
    \label{tab:mainresults}
\end{table*}

\subsection{Evaluation Metrics}
We evaluate the different approaches on three main components:
(i) \emph{safety:} how well the resulting controller $u_\theta(x, t)$ derived
from the learned value function $V_\theta(x, t)$ keeps the system safe,
(ii) \emph{level-set accuracy:} how well the (complement of) the induced BRT $\calB_\theta = \{ x \mid V_\theta(x, 0) \leq 0 \}$ captures the safe initial conditions
of the system (under the controller $u_\theta$), and
(iii) \emph{ground-truth residual error:} how close $V_\theta(x, t)$ is to the ground truth $V(x, t)$ in MSE.
Note for the subset of problems where the disturbance set is non-empty,
we utilize the disturbance $d_\theta(x, t)$ that is optimal 
for the learned value function to approximate the optimal disturbance
$d^*(x, t; u_\theta)$.

For quantifying (i) \emph{safety} and (ii) \emph{level-set accuracy}, we view $V_\theta$ as a binary classifier $\hat{h}_\theta(x) = \ind\{ V_\theta(x, 0) > 0 \}$ 
that approximates the ground truth $h^*_\theta(x) = \ind\{ J(x, 0, u_\theta, d_\theta) > 0 \}$.
For each method and problem we sample $N=10^6$ initial conditions,
count the number of true positives (TP: $\hat{h}_\theta = h^*_\theta = 1$),
false positives (FP: $\hat{h}_\theta=1, h^*_\theta=0$),
true negatives (TN: $\hat{h}_\theta = h^*_\theta = 0$), and
false negatives (FN: $\hat{h}_\theta=0, h^*_\theta=1$),
and compute the following metrics:
\begin{enumerate}[(\alph*)]
    \item \textbf{Precision:} The ratio of initial states which are correctly predicted as safe:
    \begin{align*}
        \text{Precision} =\frac{\text{TP}}{\text{TP}+\text{FN}}.
    \end{align*}
    \item \textbf{Intersection over Union (IoU):} The ratio of the intersection of predicted and actual safe sets over its union:
    \begin{align*}
        \text{IoU}=\frac{\text{TP}}{\text{TP}+\text{FP}+\text{FN}}.
    \end{align*}
    \item \textbf{Predicted Volume (PV):} 
    The percentage of initial states which are \emph{predicted} to be safe:
    \begin{align*}
        \text{PV} = \frac{\text{TP}+\text{FP}}{N}\cdot100.
    \end{align*}
    \item \textbf{True Volume (TV):} The percentage of initial states which are \emph{actually} safe under $u_\theta$:
    \begin{align*}
        \text{TV}=\frac{\text{TP}+\text{FN}}{N}\cdot 100.
    \end{align*}
\end{enumerate}
Here, $\text{TV}$ gives us a measure of the true safety of the induced controller $u_\theta$, whereas both the 
precision
and 
IoU
provide a measure of the accuracy of $\calB_\theta^c$.
In the literature,
another common metric to report is the \emph{recovered volume}~\cite{sharpless2026linearsupervision,feng2025mpcreachability,singh2025exact}, which corresponds to using split conformal prediction
to compute a threshold $\gamma$ such that the tightened classifier
$\hat{h}^\gamma_\theta(x) = \ind\{ V_\theta(x, 0) > \gamma \}$
has a false positive (FP) rate that is controlled at a fixed value
(e.g., $\text{FP} = 10^{-3}$).
To provide a more direct comparison of the various approaches, 
we choose to report
Precision, IOU, and TV metrics of \emph{pre-calibrated} $\hat{h}_\theta(x)$; 
\MethodName{} is of course compatible with any post-hoc calibration
procedure.

Finally, for evaluating (iii) \emph{ground truth residual error}
when we have access to a ground truth $V(x, t)$, 
we sample a random set of initial conditions $\{ X_0^{(i)} \}_{i=1}^{N} \stackrel{\text{i.i.d.}}{\sim} \mu_0$ drawn from a base measure $\mu_0$ (which for each problem we take to be uniform over a fixed interval for each state coordinate), and compute the
\textbf{Relative $L^2$ (RL2)} metric:
\begin{equation}
    \text{RL2}=\sqrt{\frac{\sum_{i=1}^N\left(V(X_{0}^{(i)},0)-V_{\theta}(X_{0}^{(i)},0)\right)^2}{\sum_{i=1}^N V^2(X_{0}^{(i)},0)}},
\end{equation}
where $V_\theta(x,t)$ denotes the learned value function.

\subsection{Implementation Details}

For PINNs, RAD-PINNs, and \MethodName{}, we match the model architecture and the total number of collocation points with the MPC-DeepReach models in \cite{feng2025mpcreachability,teoh2025mpcadversarial}. Therefore, we utilize a 4 layer neural network (NN) with 128 neurons per layer for the vertical drone and pursuit-evade problems and 512 neurons per layer for the F1tenth, quadrotor, and publisher-subscriber problems. All models utilize \texttt{swish} activation and boundary conditions are enforced via the parameterization~\eqref{eq:hcvalue} with quadratic constraint $\rho$ on all cases except F1tenth, where we utilize a softplus constraint $\rho$. For \MethodAbbrv{}, we use a trajectory batch size of $B_{\text{traj}}=512$ and all methods utilize a base PDE batch size of $B_{\text{pde}}=4096$. We match the total number of collocation points by adjusting the number of iterations up to $N_{\text{iter}}=400k$ iterations, at which point the PDE batch size is increased to match MPC-DeepReach; this limits the computational overhead stemming from excessive iterations and instead allows for more efficient batch computing. We note that in all cases, we utilize comparable PINNs residual collocation points instead of the MPC regression data points and trajectory rollout data points. Since MPC-DeepReach employs a MPC regression loss alongside the PINNs residual, our methods are effectively under-sampled compared to MPC-DeepReach.
All PINNs, RAD-PINNs, and \MethodAbbrv{} models, are trained using Adam with a multi-step learning-rate schedule of 
$10^{-4}$, $5 \cdot 10^{-5}$, and $10^{-5}$
for $N_{\text{traj}}/2$, $N_{\text{traj}}/4$, and $N_{\text{traj}}/4$  iterations, respectively.
We implement our algorithms in \verb|jax|~\cite{jax2018github}, and train on one NVIDIA H200 GPU using \texttt{f64} precision;
we also obtained similar results 
for \MethodAbbrv{} using \texttt{f32} precision.
For MPC-DeepReach, we utilize their open-source codebase~\cite{deepreach2024} as-is
to train models using their default architecture and algorithm hyperparameters.
All code to reproduce the following results can be found online at \href{https://github.com/sungje-park/steer2reach}{github.com/sungje-park/steer2reach}. 

\begin{figure}[tb]
    \centering
    \includegraphics[width=0.9\linewidth]{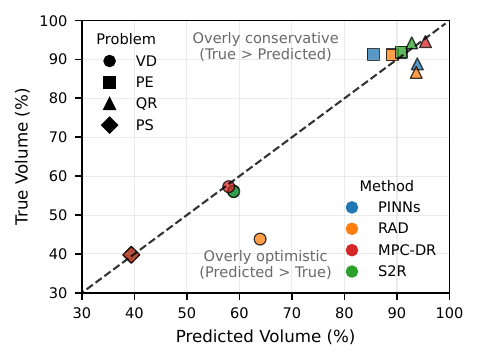}
    \caption{A plot of the predicted volume (PV) vs.\ true volume (TV) for each problem and model pair.
    The dashed black line represents the ideal 
    $\text{TV} = \text{PV}$ scaling.
    }
    \label{fig:volumeplot}
\end{figure}

\begin{figure*}[tb]
    \centering
    \begin{subfigure}[h]{.32\linewidth}
        \centering
        \includegraphics[width=1\linewidth]{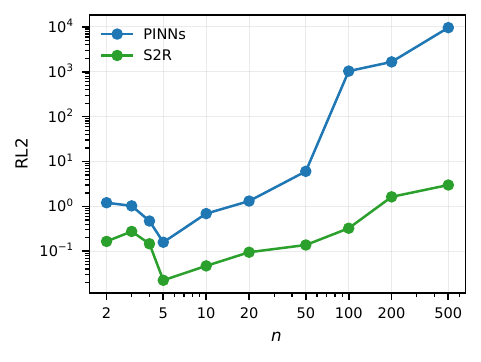}
        \caption{RL2 vs. state dimension $n$ 
            for the publisher-subscriber case.}
        \label{fig:dimrl2}
    \end{subfigure}
    \hfill
    \begin{subfigure}[h]{.32\linewidth}
        \centering
        \includegraphics[width=1\linewidth]{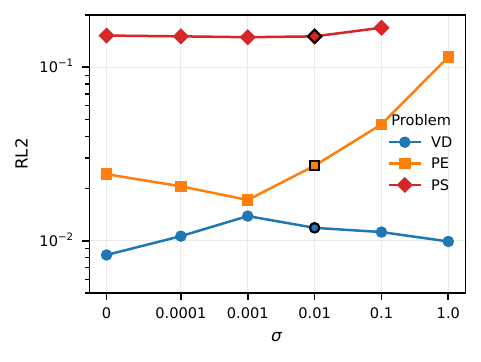}
        \caption{A plot of RL2 vs noise, $\sigma$ on problems with ground-truth solutions.}
        \label{fig:sigmarl2}
    \end{subfigure}
    \hfill
    \begin{subfigure}[h]{.32\linewidth}
        \centering
        \includegraphics[width=1\linewidth]{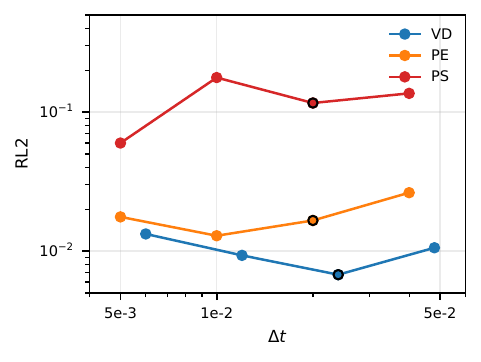}
        \caption{A plot of RL2 vs discretization fineness on problems with ground-truth solutions.}
        \label{fig:dtrl2}
    \end{subfigure}
    \caption{Ablation study plots for \MethodName{}. In (b) and (c), the outlined points represent the values used for \MethodAbbrv{} in~\Cref{tab:mainresults}.}
\end{figure*}

\subsection{Main \MethodName{} Benchmark Results}
\label{sec:results:main}

In the first set of results, we compare \MethodAbbrv{} with the three other baselines. We solve each problem using three different initialization seeds for training and report the mean and standard deviation of our metrics in \Cref{tab:mainresults}.
From \Cref{tab:mainresults}, we take away several key
trends.
First, both MPC-DeepReach and \MethodAbbrv{}
match or outperform PINNs and RAD-PINNs across most cases;
while RAD-PINNs offers some improvement
over PINNs, its performance still falls short of both MPC-DeepReach and \MethodAbbrv{} especially on the three higher dimensional problems (7D F1Tenth, 13D Quadrotor, and 40D Pub-Sub). One exception is the pursuit-evade case where RAD-PINNs achieves a better TV but a significant worse IOU and RL2 compared to MPC-DeepReach and \MethodAbbrv{}. We also see that
both MPC-DeepReach and \MethodAbbrv{}
have significantly more accurate level-sets
than PINNs and RAD-PINNs (e.g., higher IOU) even without
explicit post-hoc calibration (also see \Cref{fig:volumeplot}).

More importantly, we see that \MethodAbbrv{} is able
to be competitive with and can match the performance of MPC-DeepReach on 
most problems (with the exception of F1tenth, see 
\Cref{sec:f1tenth} for a longer discussion)
in the safety metrics (i.e.,
precision, IOU, and TV),
while achieving a lower RL2 (ground truth residual error)
by a factor of
$1.3\times$ for the pursuit-evade,
$2\times$ for the vertical drone,
and $5.7\times$ for the publisher-subscriber problem.
One notable case is the publisher-subscriber problem, where
\MethodAbbrv{} achieves the best
precision, IOU, and TV metrics among all methods,
outperforming MPC-DeepReach.
For the remaining non-F1tenth problems, \MethodAbbrv{} and MPC-DeepReach achieve comparable performance, although MPC-DeepReach sometimes has a slight advantage on safety metrics (e.g., for the quadrotor).
We hypothesize that this advantage may be partly due to its final-stage fine-tuning procedure,
which uses weighted regression to reduce false positives~\cite[Eq.\ 12]{feng2025mpcreachability}.
We emphasize that \MethodAbbrv{}'s performance
compared with MPC-DeepReach is achieved 
with a much simpler training procedure as detailed in \Cref{sec:steer2reach}---no multi-stage curriculum training pipelines and no MPC solutions are necessary.

\subsection{F1Tenth Performance}\label{sec:f1tenth}
We hypothesize that the performance discrepancy of \MethodAbbrv{} on the F1tenth problem stems from the hybrid dynamics which are piecewise nonlinear and numerically sensitive with a narrow operating region. While the low-speed kinematic dynamics (cf.~\eqref{eq:f1tenth_low_speed}) are relatively benign, the high-speed dynamics (cf.~\eqref{eq:f1tenth_high_speed}) introduces terms inversely proportional to $v$ and $v^2$ making the yaw and slip angle equations highly sensitive to discretization error. As a result, the trajectory rollouts are often unstable and produce trajectories that deviate significantly from the true system.

This sensitivity is especially problematic for \MethodAbbrv{} since the trajectory rollouts are used as a proxy for the sampling measures $\mu$. Therefore, any inaccuracies in the rollouts directly affect where the PDE residual is evaluated which can significantly degrade the quality of the training signal. In contrast, MPC-DeepReach, which also utilizes trajectory rollouts, mainly uses them to approximate regression targets which is much more tolerant to imperfect rollouts.

We investigated various approaches to limit the collocation points to the defined prescribed state domain:
\begin{itemize}
    \item \textbf{Clipping}: Collocation points exceeding the state bounds are projected back into the domain after the rollouts.
    \item \textbf{Bounding}: The state bounds are enforced throughout the rollout by projecting states back whenever they exceed the limits.
    \item \textbf{Filtering}: All rollout states that fall outside the state bounds are discarded prior to computing the loss.
\end{itemize}
However, these strategies introduced a critical sampling bias; clipping and bounding concentrate points near the limits of the admissible domain, while filtering removes later rollout states disproportionately, concentrating points around the initial portions of the trajectories. In both cases, the resulting collocation set is systematically biased and fails to provide an accurate approximation for the distribution induced by the true system dynamics.

Overall, our observations suggest that the challenge is not merely numerical instability, but the interaction between rollout errors and the induced collocation distribution.
While the proposed post-processing strategies mitigate invalid states, they do not preserve the dynamics-induced sampling distribution and therefore fail to recover training performance. Developing simple trajectory-based sampling methods that maintain both trajectory fidelity and unbiased sampling measures remains an important 
future direction.

\subsection{Scalability of \MethodName{} to High Dimensions}

Our next experiment studies the scalability of \MethodName{} as the state dimension of the problem increases. To do this, we utilize the publisher-subscriber problem, which can be adapted to any arbitrary input dimensions $n>2$ by adjusting the number of subscribers, which we vary from $n=2$ to $n=500$.
In~\Cref{fig:dimrl2}, we observe that
the performance of \MethodAbbrv{} in RL2 scales significantly better compared to
the PINNs baseline as dimensions increases.
This trend illustrates that \MethodAbbrv{}'s forward trajectory steering is able to 
effectively sampling the relevant portions of the state space for learning a 
safety value function even as the dimensionally increases.

\subsection{Ablation Studies for \MethodName{}}
\label{sec:experiments:ablation}

\begin{table}[!b]
    \centering    
    \begin{tabular}{ll|cccc}
        \toprule
         & & \multicolumn{2}{c}{PINNs} & \multicolumn{2}{c}{S2R}\\
        problem & constraint $\rho$ & TV & RL2 & TV & RL2 \\
        \midrule
        
        \multirow{5}{*}{VD 2D} & none & 46.0740 & 0.6513 & 55.8940 & 0.0150 \\
         & quadratic & \textbf{56.2202} & 0.3585 & 55.9976 & \textbf{0.0074} \\
         & softplus & 26.0273 & 0.7491 & \textbf{58.2206} & 0.0113 \\
         & swish & 38.3856 & 0.5382 & 55.8891 & 0.0160 \\
         & elu & 26.0215 & 0.7798 & 55.8472 & 0.0111 \\\midrule
        \multirow{5}{*}{PE 3D} & none & 92.1380 & \textbf{0.0865} & 92.0671 & \textbf{0.0112} \\
         & quadratic & 91.7807 & 0.1295 & 91.9540 & 0.0220 \\
         & softplus & 90.7737 & 0.1591 & 92.0171 & 0.0138 \\
         & swish & 90.8627 & 0.1213 & 92.0281 & 0.0260 \\
         & elu & 92.0754 & 0.0950 & \textbf{92.0936} & 0.0118 \\\midrule
        \multirow{5}{*}{QR 13D} & none & 85.2913 & - & 92.3224 & - \\
         & quadratic & \textbf{88.8795} & - & 94.2542 & - \\
         & softplus & 86.2966 & - & \textbf{94.2726} & - \\
         & swish & 87.3345 & - & 93.7863 & - \\
         & elu & 84.3953 & - & 93.9070 & - \\\midrule
        \multirow{2}{*}{PS 40D} & none & 0.0001 & 4.5099 & 39.7359 & 0.1472 \\
         & quadratic & 11.5185 & 2.7623 & \textbf{39.8605} & \textbf{0.1005} \\
         & softplus & \textbf{23.6384} & \textbf{2.7503} & 39.7568 & 0.1928 \\
         & swish & 21.6774 & 2.8584 & 39.7468 & 0.1915 \\
         & elu & 21.8202 & 3.1418 & 39.7333 & 0.1531 \\
        \bottomrule
        \end{tabular}

    \caption{Effect of varying the constraint function $\rho$ in \eqref{eq:hcvalue} on TV and RL2 metrics.}
    \label{tab:hardconstraint}
\end{table}

Our next set of experiments study both the hard constraint 
technique (cf.~\Cref{sec:s2rhc}), 
in addition to \MethodAbbrv{}'s performance sensitivity
to its two hyperparameters: the noise
parameter $\sigma$, and the discretization
time $\Delta t$ (cf.~\Cref{alg:fspinns}).

We first study the effect of the
hard constraints.
We test four candidate constraint functions $\rho$ for \eqref{eq:hcvalue}: 
quadratic, softplus, swish, and elu.
We also consider removing the constraint (i.e., $\rho(x) = x$).
Changing $\rho$ in 
\eqref{eq:hcvalue} does not affect the boundary constraint $V_\theta(x, T) = \ell(x)$, and hence this study 
focuses on
the benefits of strictly enforcing 
the inequality constraint $V_\theta(x, t) \leq \ell(x)$ for all time.
Note that the swish and elu functions allow some negative values and therefore do not strictly enforce the inequality constraint; however, both functions
are bounded below and hence restrict
how much $V_\theta(x, t)$ can cross above $\ell(x)$.
The results are shown in 
\Cref{tab:hardconstraint}.
Here, we see that hard enforcement of the inequality constraint can lead to improved performance (measured by both TV and RL2) 
compared to the unconstrained parameterization, for both PINNs and \MethodAbbrv{}.
However, there is not a clear trend as to which choice of $\rho$ is the best, 
and more importantly the gains in \Cref{tab:hardconstraint} are considerably
less than those in \Cref{tab:mainresults} (which fixes a quadratic $\rho$) between different algorithms,
further illustrating that the forward trajectory steering is the key component.

We next turn to studying the influence of \MethodAbbrv{}'s two hyperparameters 
on its performance. \Cref{fig:sigmarl2} illustrates the effect of various values of the noise parameter $\sigma$ on
RL2 error, and \Cref{fig:dtrl2} showcases the effect of various discretization times
$\Delta t$ also on RL2 error. Regarding the noise $\sigma$,
\Cref{fig:sigmarl2} shows that RL2 error can be improved by tuning the noise,
illustrating that the exploration induced by the forward trajectory SDE
can be helpful for learning. However, the best choice of $\sigma$ is heavily
problem dependent. We emphasize that in our main experiments in \Cref{sec:results:main},
we do \emph{not} tune $\sigma$ for each problem setting, but instead
we fix $\sigma=0.01$ across all instances.
Next, regarding the discretization time $\Delta t$, in
\Cref{fig:dtrl2} we see that $\Delta t$ also impacts RL2 error in a problem-dependent
way. Interestingly, we see that smaller $\Delta t$ does not necessarily lead to 
lower error despite more accurate SDE integration, 
a trend we leave investigating further to future work.
As with $\sigma$, we do not tune $\Delta t$ for each problem setting in \Cref{tab:mainresults}, 
but instead we fix $\Delta t$ so that $N_{\text{traj}} = 50$ for each problem
(cf.~\Cref{alg:fspinns}).

\section{Conclusion and Future Work}
\label{sec:conclusion}

We propose \MethodName{}, a simple HJ reachability PINNs-based solver 
that utilizes forward SDE trajectory steering to adaptively sample
collocation points. Through comparisons with various PINNs-based methods
including SoTA methods such as MPC-DeepReach, 
we demonstrate that \MethodAbbrv{} achieves 
competitive safety performance
and lower relative $L_2$ error compared to 
SoTA solvers, while being much simpler to implement and not requiring
multi-stage training pipelines or auxiliary semi-supervision.

This work opens up some exciting future directions, for which we list a couple.
First, while \MethodAbbrv{} is inspired by reflected BSDE algorithms and borrows 
its forward SDE sampling from BSDE methods, it is still a PINNs method that minimizes
PDE residual. Designing reflected BSDE solvers that yield even higher quality
safety value functions than \MethodAbbrv{} is hence an interesting future direction.
Second, applying \MethodAbbrv{} to the visuomotor control setting 
for enabling scalable and accurate latent reachability analysis~\cite{nakamura2025generalizing}
is also another important direction.

\section*{Acknowledgments}

Anthropic's Claude Code~\cite{anthropic2026code} and OpenAI's Codex~\cite{openai2026codex} assisted with implementation, testing, and analysis of the experimental code. All algorithmic design, numerical methods, and experimental validations were performed by the authors.
OpenAI's ChatGPT~\cite{openai2026chat} assisted with writing portions of the manuscript, in addition to finding relevant literature
and instantiating reflected BSDEs for HJI-VI problems.

\bibliographystyle{IEEEtran}
\bibliography{paper}

\appendix
\allowdisplaybreaks
\section{Appendix}

\subsection{RAD Algorithm}\label{sec:appendix:radalgo}
\begin{algorithm}[h]
    \caption{RAD Sampling~\cite{wu2023comprehensiveadaptive}}
    \label{alg:rad}
    \begin{algorithmic}
        \small
        \setlength{\baselineskip}{1.3\baselineskip}
        \Input $V_{\theta}(x,t)$, $B_{\text{pde}}$, $N_{\text{cand}}$, $k$, $c$
        \Output  $\{(x_b,t_b)\}_{b=1}^{B_{\text{pde}}}$
        \State $(x_i,t_i)\sim U(\Omega\times[t_0,t_f]),\quad i=1,\dots,N_{\text{cand}}$
        \State $r_i=\left|R_{V_\theta}(x_i,t_i)\right|$
        \State $q_i=(r_i)^k$
        \State $\bar q=\tfrac{1}{N_{\text{cand}}}\sum_{m=1}^{N_{\text{cand}}} q_m$
        \State $w_i=\tfrac{q_i}{\bar q}+c$
        \State $p_i=\tfrac{w_i}{\sum_{j=1}^{N_{\text{cand}}} w_j}$
        \State $\{i_b\}_{b=1}^{B_{\text{pde}}} \sim \text{SampleWithoutReplacement}((p_1,\dots,p_{N_{\text{cand}}}), B_{\text{pde}})$
        \State $(x_b,t_b)=(x_{i_b},t_{i_b}),\quad b=1,\dots,B_{\text{pde}}$
    \end{algorithmic}
\end{algorithm}

\subsection{Problem Dynamics}\label{appendix:sec:problems}
In this section, we describe the five problems used in this work. The vertical drone, f1tenth, quadrotor, and publisher-subscriber cases are considered in~\cite{feng2025mpcreachability} while the pursuit-evade case is considered in~\cite{bansal2020deepreach}. 
Unless stated otherwise, all cases are considered over the time horizon $t\in[0,1]$. 

\noindent\textbf{2D Vertical Drone:}
We consider a drone moving vertically along the z-axis in time horizon $t\in[0,1.2] $ with dynamics:
\begin{equation}
    x=\mkmat{z\\v_z},\quad \dot x=\mkmat{v_z\\Ku-g}
\end{equation}
where $z\in[-.5,3.5]$ denotes the height, $v_z\in[-4,4]$ denotes vertical velocity, $g=9.8$ is gravitational acceleration, and $K=12$ is the constant control gain parameter. The control input $u\in[-1,1]$ represents the vertical acceleration effort. The failure set is given as,
\begin{equation}
    \calL=\set{x \mid |z-1.5|\geq 1.5}
\end{equation}
which represents the drone hitting the ground or ceiling.

\noindent\textbf{3D Pursuit-Evade:}
We consider a collision avoidance problem between two vehicles with dynamics,
\begin{align*}
    x=\mkmat{x_1\\x_2\\x_3},\quad \dot x=\mkmat{-v_e+v_p\cos x_3+\omega_e x_2\\v_p\sin x_3-\omega_ex_1\\w_p-w_e}
\end{align*}
where $x_1,x_2\in[-1,1]$ denotes the relative position between vehicles, $x_3\in[-\pi,\pi]$ denotes the relative heading, and $v_e=v_p=.75$ are the linear velocity of the evader and pursuer. The control and disturbance $\omega_e,\omega_p\in[-3,3]$ represent the angular velocities of the evader and pursuer. The failure set for the problem is given as,
\begin{equation}
    \calL=\set{x \mid \norm{(x_1,x_2)} \leq .25}
\end{equation}
which represents the vehicles colliding.

\noindent\textbf{7D F1tenth:}
We consider a hybrid vehicle dynamics system with a low-speed kinematic mode and a high-speed dynamic mode over the time horizon $t\in[0,1]$. The states for this system are given as $x=\mkmat{p_x,p_y,\phi,v,\theta_{\text{yaw}},\omega_{\text{yaw}},\beta_{\text{slip}}}$ with hybrid dynamics,
\begin{equation}
    \dot{x}=
    \begin{cases}
        f_{\mathrm{kin}}(x,u), & |v|<.5,\\
        f_{\mathrm{dyn}}(x,u), & |v|\geq .5,
    \end{cases}
\end{equation}
where the low-speed kinematic dynamics are
\begin{equation}
    f_{\mathrm{kin}}(x,u)=
    \mkmat{
        v\cos(\theta_{\text{yaw}})\\
        v\sin(\theta_{\text{yaw}})\\
        \dot{\phi}\\
        a\\
        \dfrac{v}{l_r+l_f}\tan(\phi)\\
        \dfrac{a}{l_r+l_f}\tan(\phi)
        +\dfrac{v}{(l_r+l_f)\cos^2(\phi)}\dot{\phi}\\
        0
    }, \label{eq:f1tenth_low_speed}
\end{equation}
and the high-speed dynamic dynamics are
\begin{equation}
    f_{\mathrm{dyn}}(x,u)=
    \mkmat{
        v\cos(\theta_{\text{yaw}}+\beta_{\text{slip}})\\
        v\sin(\theta_{\text{yaw}}+\beta_{\text{slip}})\\
        \dot{\phi}\\
        a\\
        \omega_{\text{yaw}}\\
        -\dfrac{\mu m}{vI(l_r+l_f)}
        (
            l_f^2C_{S_f}(gl_r-ah)\\
            +l_r^2C_{S_r}(gl_f+ah)
        )\omega_{\text{yaw}}\\
        +\dfrac{\mu m}{I(l_r+l_f)}
        (
            l_rC_{S_r}(gl_f+ah)\\
            -l_fC_{S_f}(gl_r-ah)
        )\beta_{\text{slip}}\\
        +\dfrac{\mu m}{I(l_r+l_f)}
        l_fC_{S_f}(gl_r-ah)\phi,\\
        \Big[
            \dfrac{\mu}{v^2(l_r+l_f)}
            (
                l_rC_{S_r}(gl_f+ah)\\
                -l_fC_{S_f}(gl_r-ah)
            )-1
        \Big]\omega_{\text{yaw}}\\
        -\dfrac{\mu}{v(l_r+l_f)}
        (
            C_{S_r}(gl_f+ah)\\
            +C_{S_f}(gl_r-ah)
        )\beta_{\text{slip}}\\
        +\dfrac{\mu}{v(l_r+l_f)}
        C_{S_f}(gl_r-ah)\phi
    }. \label{eq:f1tenth_high_speed}
\end{equation}
Here, $p_x\in[0,62.5]$ and $p_y\in[0,50]$ denote the global position of the vehicle, $\phi\in[-.4189,.4189]$ denotes the front-wheel steering angle, $v\in[0,8]$ denotes the vehicle speed, $\theta_{\text{yaw}}\in[-\pi,\pi]$ denotes the yaw angle, $\omega_{\text{yaw}}\in[-5,5]$ denotes the yaw rate, and $\beta_{\text{slip}}\in[-.8,.8]$ denotes the slip angle at the vehicle center. The control inputs are defined as $u=\mkmat{\dot{\phi}\\a}\in[-3.2,3.2]\times[-9.51,9.51]$, where $\dot{\phi}$ represents the steering rate and $a$ represents the longitudinal acceleration. The longitudinal acceleration is further capped at high vehicle speeds. The vehicle parameters are defined as $\mu=1.0489$, $C_{S_f}=4.718$, $C_{S_r}=5.4562$, $l_f=.15875$, $l_r=.17145$, $h=.074$, $m=3.74$, $I=.04712$, and $g=9.81$, representing the surface friction coefficient, front and rear cornering stiffness coefficients, distances from the center of gravity to the front and rear axles, center-of-gravity height, vehicle mass, moment of inertia about the vertical axis, and gravitational acceleration, respectively. The failure set for this system is defined as
\begin{equation}
    \calL=\set{x\mid l_{\mathrm{track}}(p_x,p_y)\leq 0},
\end{equation}
where $l_{\mathrm{track}}(p_x,p_y)$ denotes the signed distance to the edge of the track. In this problem, the failure set represents the vehicle leaving the track and colliding with the curbs.

\noindent\textbf{13D Quadrotor:}
We consider a quadrotor system with states $x=\mkmat{p_x,p_y,p_z,q_\omega,q_x,q_y,q_z,v_x,v_y,v_z,\omega_x,\omega_y,\omega_z}$ and dynamics,
\begin{equation}
    \dot x=\mkmat{v_x\\
        v_y\\
        v_z\\
        -(\omega_x\cdot q_x)/2-(\omega_y \cdot q_y)/2-(\omega_z\cdot q_z)/2\\
        (\omega_x\cdot q_\omega)/2+(\omega_z\cdot q_y)/2-(\omega_y\cdot q_z)/2\\
        (\omega_y\cdot q_\omega)/2-(\omega_z\cdot q_x)/2+(\omega_x\cdot q_z)/2\\
        (\omega_z\cdot q_\omega)/2+(\omega_y\cdot q_x)/2-(\omega_x\cdot q_y)/2\\
        CT\cdot(2\cdot q_\omega\cdot q_y+2\cdot q_x\cdot q_z)F/m\\
        CT\cdot(-2\cdot q_\omega\cdot q_x+2\cdot q_y\cdot q_z)F/m\\
        Gz-CT\cdot(2\cdot q_x^2+2\cdot q_y^2-1)F/m\\
        \alpha_x-\frac 59 \omega_y\cdot \omega_z\\
        \alpha_y+\frac 59 \omega_x\cdot \omega_z\\
        \alpha_z}
\end{equation}
where $p_x,p_y,p_z\in[-3,3]$ denotes the position, $v_x,v_y,v_z\in[-5,5]$ denotes the linear velocities, $q_\omega,q_x,q_y,q_z\in[-1,1]$ denotes the quaternion, and $\omega_x,\omega_y,\omega_z\in[-5,5]$ denotes the angular velocities. The control inputs are defined as $F,\alpha_x,\alpha_y,\alpha_z\in[-20,20]\times [-8,8]^2\times [-4,4]$. $CT=1$ represents the lifting coefficient, $m=1$ represents the mass, and $Gz=-9.81$ represents gravitational acceleration. The failure set for this system is defined as,
\begin{align*}
    \mathbf{v}^n&=\mathbf{q}\otimes \mathbf{e}_3^q\otimes \bar{\mathbf{q}}\\
    d_x&=\frac{r_a^2p_x^2\nu_z^2}{p_x^2\nu_x^2+p_x^2\nu_z^2+2p_xp_y\nu_x\nu_y+p_y^2\nu_y^2+p_y^2\nu_z^2}\\
    d_y&=\frac{r_a^2p_y^2\nu_z^2}{p_x^2\nu_x^2+p_x^2\nu_z^2+2p_xp_y\nu_x\nu_y+p_y^2\nu_y^2+p_y^2\nu_z^2}\\
    l(x)&=\max\left(\sqrt{x^2+y^2}-\sqrt{d_x+d_y},0\right)-r_0\\
    \calL&=\set{x \mid l(x)\leq 0}
\end{align*}
where $\mathbf{q}=q_\omega + q_x\cdot i+q_y\cdot j+q_z\cdot k$, $\mathbf{\bar q}=q_\omega - q_x\cdot i-q_y\cdot j-q_z\cdot k$, $\mathbf{e}_3^q=0+0i+0j+1k$, $r_a=.17$ represents the radius of the quadrotor, and $r_0=.5$ represents the radius of the obstacle. In this problem, the failure set represents a collision between the quadrotor and a cylindrical obstacle with infinite length in the z axis.

\noindent\textbf{40D Publisher-Subscriber:}
Finally, we consider a publisher-subscriber system with dynamics,
\begin{equation}
    \mkmat{\dot x_0\\\dot x_i}=\mkmat{a&0\\-1&a}\mkmat{x_0\\x_i}+\mkmat{0\\b}u_i+\mkmat{\alpha \sin(x_0)x_0^2\\-\beta x_0^2x_i}
\end{equation}
where $x_0$ represents the publisher state and $x_i$ represents each subscriber state. We define $a=-.5$, $b=.4$, $\alpha=0$, and $\beta=20$ and constrain the input to $u_i\in[-.5,5]$. This problem defines a target set as,
\begin{equation*}
    \calL=\Set{x \mid \frac{1}{2}(x_0^2+x_i^2-.5)\leq 0,\,\,\forall i=1,\dots,D-1},
\end{equation*}
where $D$ represents the number of dimensions.

\end{document}